\documentclass[a4paper]{jpconf}
\usepackage{graphicx}
\usepackage{epstopdf}
\usepackage{color}

\newcommand{\msol}{\,$M_{\odot}$}
\newcommand{\mhz}{\,$\mu$Hz}

\begin{document}
\title{Variation in the frequency separations with activity and impact on stellar parameter determination}
\author{O L Creevey$^{1,2}$, D Salabert$^{1,2}$, R A Garc\'\i a$^3$}

\address{$^1$ Instituto de Astrof\'isica de Canarias (IAC), E-38200 La Laguna, Tenerife, Spain}
\address{$^2$ Dept. de Astrof\'isica, Universidad de La Laguna (ULL), E-38206 La Laguna, Tenerife, Spain}
\address{$^3$ Laboratoire AIM, CEA/DSM-CNRS, Universit\'e Paris 7 Diderot, IRFU/SAp, Centre de Saclay, F-91191 Gif-sur-Yvette, France}
%\address{$^3$ Laboratoire AIM, CEA/DSM-CNRS-Universit\'e Paris Diderot, CEA, IRFU, SAp, F-91191, Gif-sur-Yvette, France}

\ead{orlagh@iac.es, salabert@iac.es, rgarcia@cea.fr}

%\author{Jacky Mucklow}

%\address{Production Editor, \jpcs, \iopp, Dirac House, Temple Back, Bristol BS1~6BE, UK}

%\ead{jacky.mucklow@iop.org}

\begin{abstract}
Frequency separations used to infer global properties of stars through asteroseismology can change depending on the strength and at what epoch of the stellar cycle the p-mode frequencies are measured.
In the Sun these variations have been seen, even though the Sun is a {low-activity star}.
%\red{\bf I do not like inactive because we know that the frequencies change, etc}
In this paper, we discuss these variations and their impact on the determination 
of the stellar parameters (radius, mass and age) for the Sun.
Using the data from maximum and minimum activity, we fitted an age for the Sun that differs on average by 0.2 Gyr: slightly older during minimum activity.
The fitted radius is also lower by about 0.5\% for the solar effective temperature during minimum.
\end{abstract}

\section{Introduction}
%\red{TO BE FILLED IN: Rafa, can you do this?}
%\red{You can use the command "msol" or  "mhz" for solar mass and microHz}
The p-mode oscillation frequencies vary with the activity cycle of the Sun \cite{sal09} and other solar-like stars \cite{gar10}. However, as observed in the Sun, these temporal variations present differences in amplitude, phase, and frequency among individual modes. Thus the frequency separations computed will be observed to vary with time.  
Comparing these frequency separations with those from stellar models will lead to 
differences in the fitted global parameters (radius, mass, age) of the star.
Generally, when the frequencies are measured we do not know the phase
of the activity cycle, unless the star can be measured continuously for 
long periods of time.
We should then take into account when inferring the stellar parameters, that
these values may depend on when the data was obtained.

\section{Analysis of time-series observations and determination of frequencies}
We analyzed 5202 days of {velocity} observations collected by the space-based instrument GOLF onboard {\it SoHO} spacecraft, covering more than 14 years between 1996 and 2010 with an overall duty cycle of 95.4\% \cite{gar05}. {This dataset was split into non-independent, contiguous 365-day subseries with 91.25-day overlap} and their associated power spectra fitted to extract the mode parameters \cite{sal07} using a standard likelihood maximization function (power spectrum with a $\chi^2$ with 2 d.o.f. statistic). The formal uncertainties in each parameter were then derived from the inverse Hessian matrix. 

\begin{figure*}
\includegraphics[width = 0.5\textwidth]{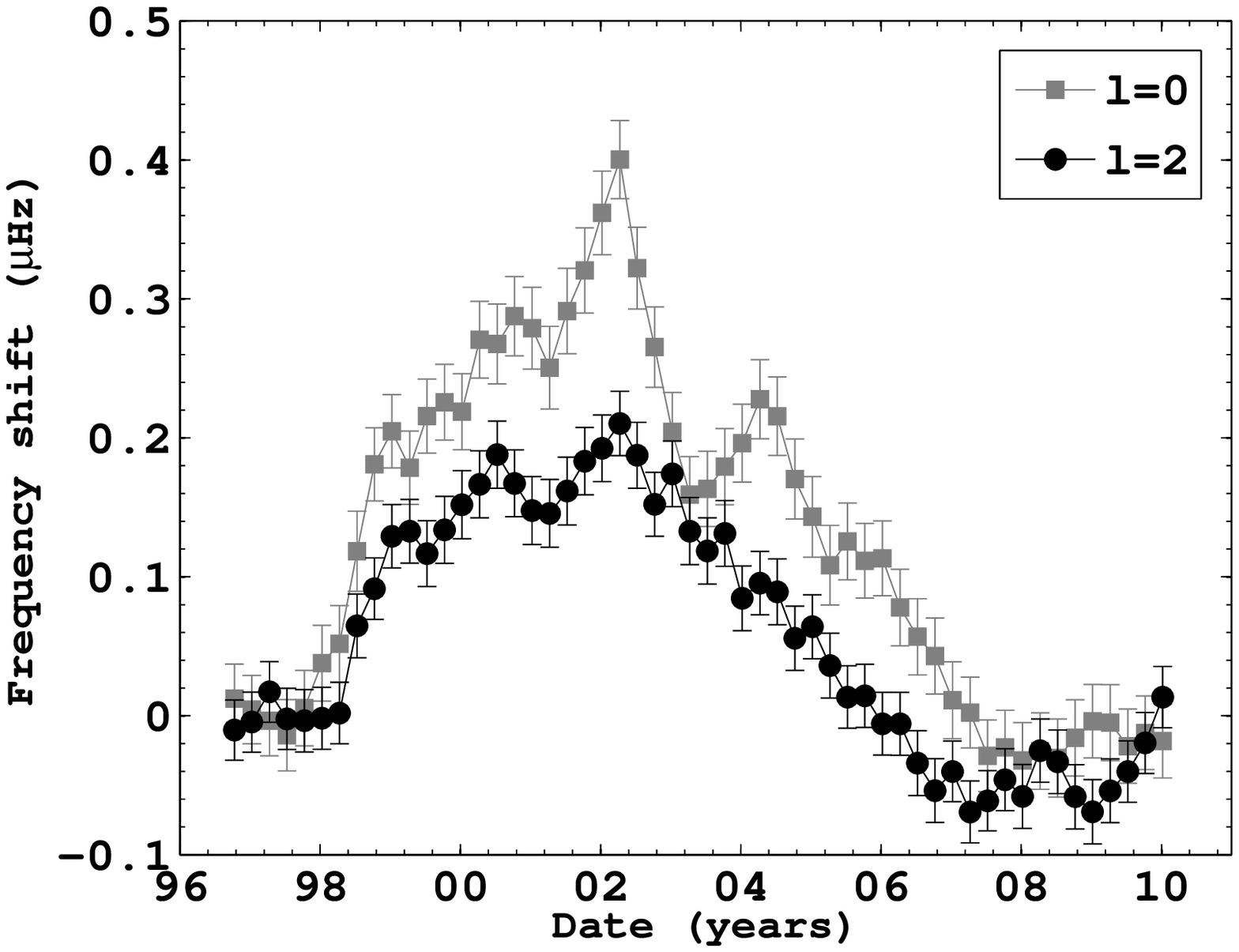}
\includegraphics[width = 0.5\textwidth]{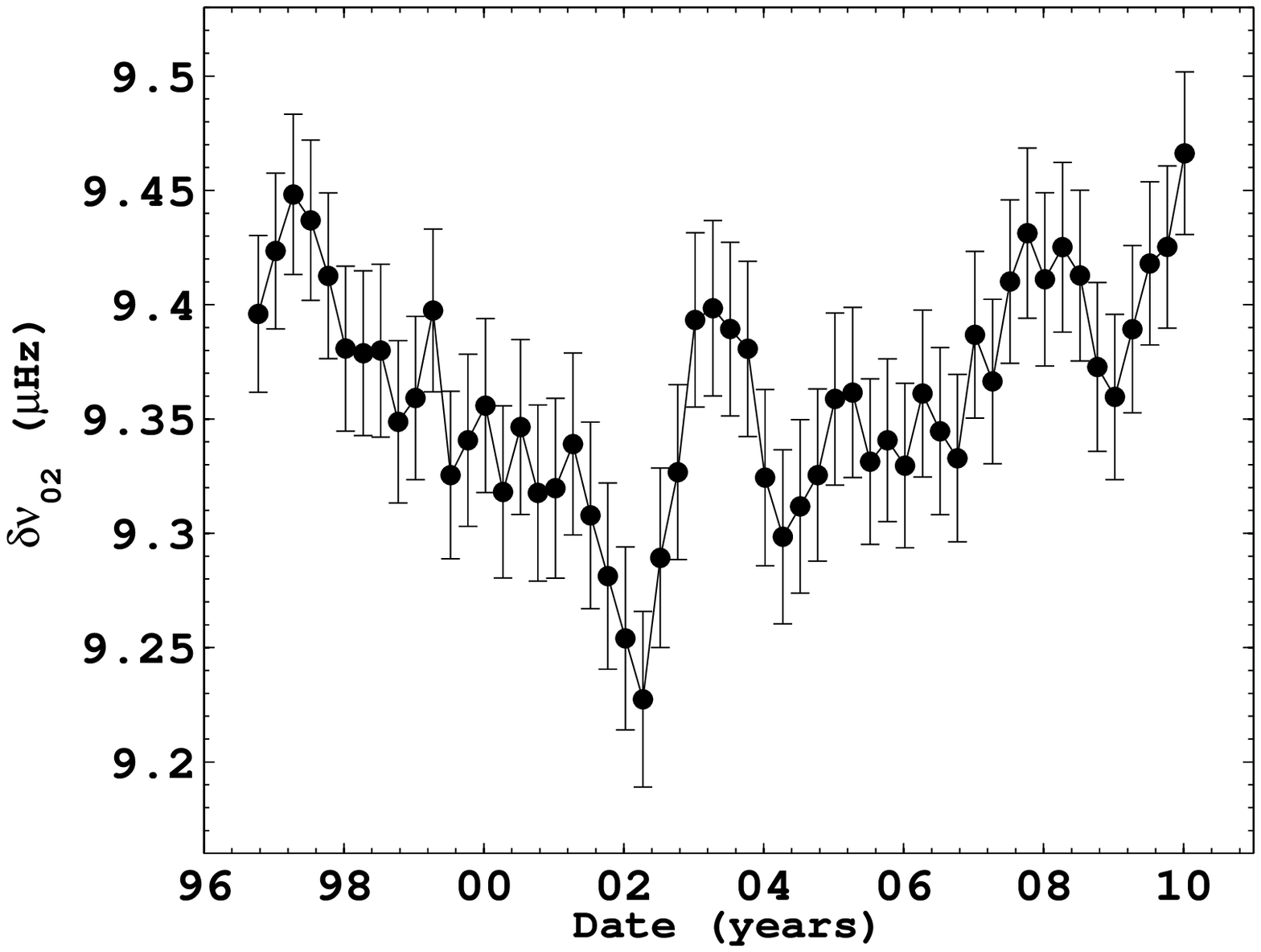}
\caption{\label{fig:data} {Left: Frequency shifts ($\mu$Hz) of the $l = 0$ ($\fullcircle$) and $l = 2$ ($\fullsquare$) modes measured from GOLF data. Right: Temporal variations of the averaged small frequency separation $\delta\nu_{02}$ ($\mu$Hz). }}
\end{figure*}

The individual frequency separations measured with 365-day time series were obtained with a precision of about 0.07~$\mu$Hz.
Figure~\ref{fig:data} (left panel) shows the temporal variations of the individual $l = 0$ and $l = 2$ mode frequencies averaged between 2200 and 3300~$\mu$Hz (the reference values being taken as the average over 1996--1997). It is clear that the individual Sun-as-a-star p-mode frequencies have different temporal variations \cite{sal09}, which are consistent between radial velocity (GOLF) and intensity VIRGO observations \cite{sal10}. The averaged large ($\Delta\nu$) and small ($\delta\nu$) frequency separations also show significant temporal variations with solar activity. For example, Figure~\ref{fig:data} (right panel) shows that the small frequency separation $\delta\nu_{02}$ varies from peak-to-peak by about $0.2 \pm 0.02 \mu$Hz over the solar cycle, {which is presumably consistent with being signatures from surface effects}. For a broader perspective, it is very important to note that the Sun is considered a low-activity star, and many solar-type stars could exhibit much larger variations with activity cycle. 

\section{Solar global parameters}
\subsection{Fitting strategy}
In order to determine the global model parameters of the Sun $P$, 
we match the observational data $O$ to 
the observables $M$ from stellar models \cite{jcd08a,jcd08b} which are characterized by $P$.  
$P$ comprises the mass M, age A, initial hydrogen (or helium) mass fraction $X_0$, initial heavy element mass fraction $Z_0$, and the mixing-length parameter $\alpha$ (from the standard mixing-length theory of \cite{bv54}).
The parameters that best describe the data are obtained by minimizing a $\chi^2$ function:
\begin{equation}
\chi^2 =  \sum_{i=1}^M  \left (\frac{O_i - M_i}{\sigma_i} \right )^2,
\label{eqn:chi2}
\end{equation}
where there are $i=1,2,...,M$ independent observations and $\sigma$ is the observational error.
We use the Levenberg-Marquardt algorithm (LM) to minimize Eq.~\ref{eqn:chi2} 
\cite{cre07,met09,ste09,cha10,mat10}.
%Creevey et al. 2007, Metcalfe et al. 2009, Stello et al. 2009, Mathur et al 2010, Chaplin et al. 2010).

The global objective of this work is to investigate the effect of the shifted frequency values on the determination of  $P$ for any star with solar-like oscillations.
So we must define a coherent and consistent 
method to fit the observational data to obtain
the set of best-fitting parameters not only for the Sun, but for other stars too. 
Our strategy is to first fit the non-seismic and the average seismic quantities to obtain $P$ to 
a reasonable range, and then using $P$ as initial guesses, proceed to use the individual
frequency separations (large and small) to obtain the final set of best-fitting parameters.

LM is a local minimization method and can therefore be sensitive to initial conditions.  
To ensure that we obtain a global minimum we search for the best parameters by 
initializing the minimization with different parameter values.  
We set the initial $M$ as 1.0 \msol, and the initial A varies between 4 and 6 Gyr in steps of 0.5 Gyr.
Additionally, we hold the parameter $X_0$ fixed during each minimization, and instead
repeat the process for three values of $X_0$: 0.69, 0.71, 0.73.  This gives in total 15 minimizations using the 
input data at maximum activity and again at minimum activity.

\subsection{Results}

\begin{figure}
\includegraphics[width = 0.5\textwidth]{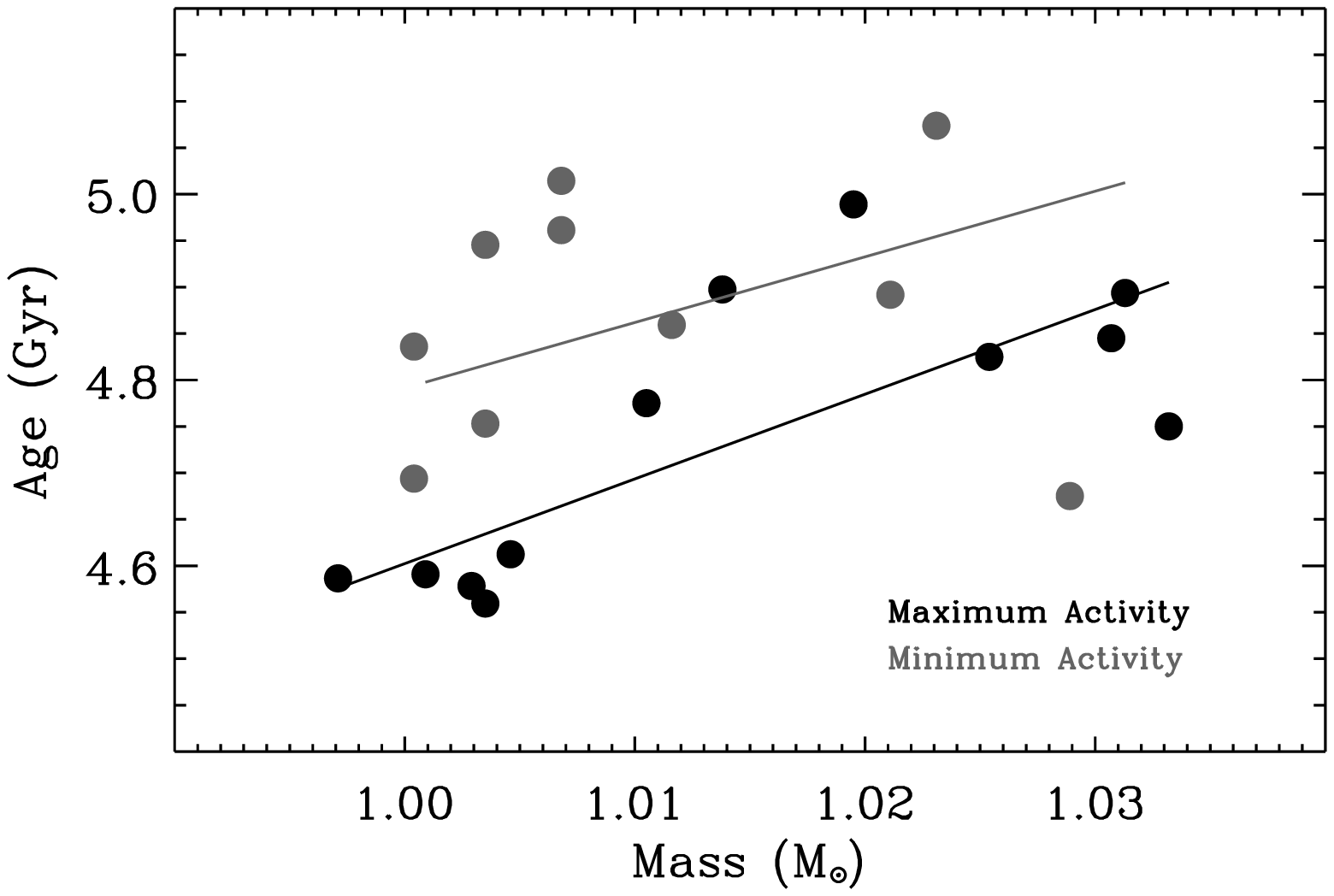}
\includegraphics[width = 0.5\textwidth]{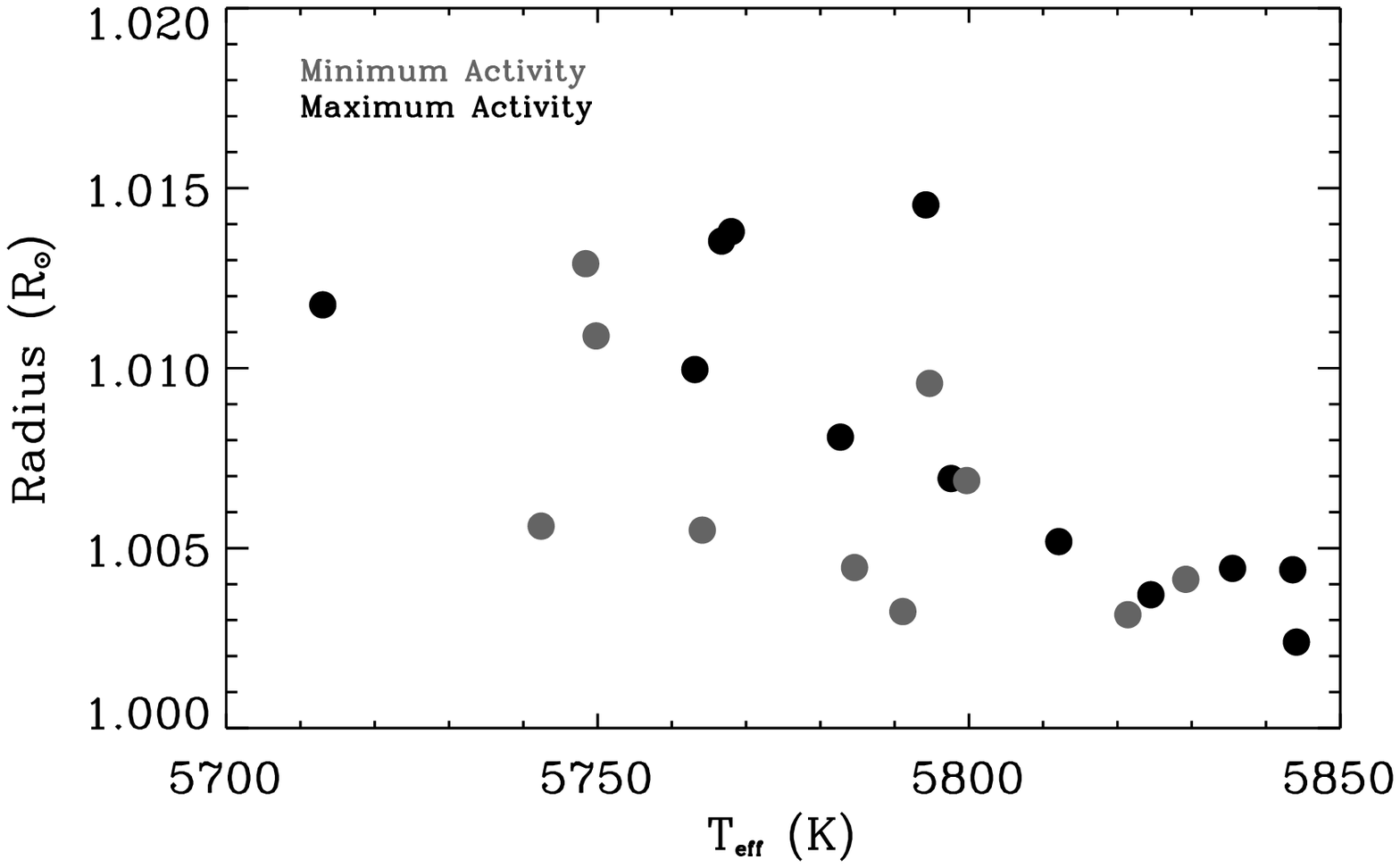}
\caption{Left: Fitted age versus mass using the individual large and small frequency separations calculated at maximum (black) and minimum (grey) activity.  Each point corresponds to a minimization. The lines are linear fits to the results.  Right: The fitted radius and effective temperature for the same minimizations.
\label{fig:models}
}
\end{figure}

From the 15 best-fitted $P$, we selected those sets where $\chi^2 \le 29.1$ (4 fitted parameters, 19 data points to give 14 degrees of freedom at the 1\% confidence level) and where the fitted $T_{\rm eff}$ fell to 
within 1$\sigma$ of 5777 K ($T_{\rm eff}$ was not a constraint in the second part of the fitting strategy).  
These sets are represented in Fig.~\ref{fig:models}, where we show the values of age versus mass on the left panel, and radius versus effective temperature on the right.
The dark/lighter filled circles are the results for fitting during maximum/minimum activity.

Fig.~\ref{fig:models} left panel shows an offset in the fitted $P$ between 
minimum and maximum activity.  
To quantify this offset, we fitted a linear function to each group of results.  
The difference between the fitted ages at opposite activity phases is 0.2 Gyr.  
%This value has 
Alternatively if we take 4.8 Gyr as the solar age, the mass is fitted as a 1.0 \msol\ star  at
minimum, and 1.03 \msol\ at maximum.

The right panel of Fig.~\ref{fig:models} shows the fitted $R$ and $T_{\rm eff}$ for these same models
(neither were input constraints). 
We find that at  minimum activity, the Sun's $T_{\rm eff}$ is about 50 K cooler for a given $R$, or
about 0.5\% smaller in $R$ for a given $T_{\rm eff}$.
%Note that neither of these were used as constraints while fitting the individual frequency separations.
%The solar parameters at minimum activity tend to be about 50 K cooler for a given radius, or
%about 0.5\% smaller in radius for a given effective temperature.

\section{Conclusions}
%\begin{itemize}
Measuring the oscillation frequencies during different phases of a stellar activity cycle will lead to different values of the individual frequencies. Thus, estimates of the large and small frequency separations will be different depending on the observing period.
Using 365-day subseries of GOLF data, we determined the individual frequency separations to about 0.07 \mhz.

%\item 
We used stellar models to determine the best-fitting parameters using the observations taken at minimum and maximum activity.  We found that the age of the star is on average 0.2 Gyr older using the values from minimum activity, or for a given age, the mass is about 2\% larger using the data at 
maximum activity. 
We also found a small discrepancy in the fitted radius and effective temperature.  At minimum activity, 
the Sun is about 50 K cooler for a given fitted radius, or the radius is about 0.5\% smaller for a
given effective temperature.

Although we still need to study whether these differences in fitted values are detectable given the expected uncertainties, we note that \cite{jcd10} quote an error in the age of the planet-hosting star HAT-P-7 of 0.26 Gyr, of the order of this detected change.
The Sun, however, is  considered a low-activity star, and 
much stronger activity cycles on other solar-type stars have already been detected \cite{gar10}.
With long time-series data such as those from CoRoT or Kepler, we will not only detect
stellar activity cycles, but we will obtain very high precision on the seismic data, that the 
uncertainties in the fitted parameters will be smaller than the fitted changes in these parameters.

% We thus expect to measure a larger difference in seismic quantities, such as HD~49933 \red{references here}.

%We still need to study whether these differences in fitted values are detectable considering the expected uncertainties.  Observing some stars over a long period of time (such as with Kepler) will allow us to reach extreme precisions that the differences will be noticeable.

%\item The Sun is not considered an active star, but many other solar-type stars will be more active.  We thus expect to measure a larger difference in seismic quantities, such as HD~49933 \red{references here}.

%\end{itemize}
%Recently CD et al quoted an error in the age of the planet-hosting star HAT-P-7 of 0.26 Gyr

\ack
The authors want to thank Catherine Renaud for the calibration and preparation of the GOLF dataset. The GOLF instrument onboard SoHO is a cooperative effort of many individuals, to whom we are indebted. SoHO is a project of international collaboration between ESA and NASA.This research was in part supported by the European Helio- and Asteroseismology Network (HELAS), a major international collaboration funded by the European Commission's Sixth Framework Programme. DS acknowledges the support of the grant PNAyA2007-62650 from the Spanish National Research Plan. This work has been partially supported by the CNES/GOLF grant at the SAp CEA-Saclay.

\end{document}